\documentclass[10pt, preprint2]{emulateapj}
\usepackage{verbatim}

\newcommand{\target}{LHS\,3844}
\newcommand{\planet}{LHS\,3844\,b}

\newcommand{\rsun}{\ensuremath{R_\sun}}
\newcommand{\msun}{\ensuremath{M_\sun}}
\newcommand{\lsun}{\ensuremath{L_\sun}}

\newcommand{\rearth}{\ensuremath{R_\earth}}

\newcommand{\mjup}{\ensuremath{M_{\rm Jup}}}

\newcommand{\teff}{\ensuremath{T_{\rm eff}}}

\newcommand{\rpl}{\ensuremath{R_{p}}}

\newcommand{\kms}{\ensuremath{\rm km\,s^{-1}}}
\newcommand{\ms}{\ensuremath{\rm m\,s^{-1}}}

\newcommand{\rstar}{\ensuremath{R_\star}}
\newcommand{\mstar}{\ensuremath{M_\star}}
\newcommand{\lstar}{\ensuremath{L_\star}}

\newcommand{\teffstar}{\ensuremath{T_{\rm eff\star}}}
\newcommand{\rhostar}{\ensuremath{\rho_\star}}
\newcommand{\loggstar}{\ensuremath{\log{g_{\star}}}}

\newcommand{\arstar}{\ensuremath{a/\rstar}}

\newcommand{\Kepler}{{\it Kepler}}

\newcommand{\tess}{{\it TESS}}

\newcommand{\Lc}{Light curve\ }

\newcommand{\gcmc}{\ensuremath{\rm g\,cm^{-3}}}
\newcommand{\ergscmsq}{\ensuremath{\rm erg\,s^{-1}\,cm^{-2}}}

\newcommand{\tessfitP}{\ensuremath{0.46292792\pm0.0000016}}
\newcommand{\tessfitTc}{\ensuremath{1325.72568\pm0.00025}}
\newcommand{\tessfitRratio}{\ensuremath{0.0640\pm0.0007}}
\newcommand{\tessfitb}{\ensuremath{0.221\pm0.038}}
\newcommand{\tessfitaor}{\ensuremath{7.1059\pm0.028}}
\newcommand{\tessfitinc}{\ensuremath{88.22\pm0.30}}
\newcommand{\tessfitRp}{\ensuremath{1.32\pm0.02}}
\newcommand{\tessfitsemi}{\ensuremath{0.00623\pm0.00015}}
\newcommand{\tessfitTdur}{\ensuremath{31.08\pm0.25}}
\newcommand{\tessfitTeq}{\ensuremath{805\pm20}}
\newcommand{\tessfitFinso}{\ensuremath{0.0954\pm0.00070}}
\newcommand{\tessfitdistance}{\ensuremath{14.9\pm0.01}}
\newcommand{\tessfitTingress}{\ensuremath{3.73_{-0.7}^{+0.4}}}
\newcommand{\tessfitrhostar}{\ensuremath{31.69\pm0.37}}

\shorttitle{\tess\ Discovery of a Planet Around LHS\,3844}
\shortauthors{}

\begin{document}

\title{\tess\ Discovery of an Ultra-Short-Period Planet Around the Nearby M Dwarf LHS\,3844}

\author{
Roland K.\ Vanderspek\altaffilmark{1},
Chelsea X.\ Huang\altaffilmark{1,2},
Andrew Vanderburg\altaffilmark{3},
George R.\ Ricker\altaffilmark{1},
David W.\ Latham\altaffilmark{4},
Sara Seager\altaffilmark{1,17},
Joshua N.\ Winn\altaffilmark{5},
Jon M.\ Jenkins\altaffilmark{4},
Jennifer Burt\altaffilmark{1,2},
Jason Dittmann\altaffilmark{1,17},
Elisabeth Newton\altaffilmark{1},
Samuel N.\ Quinn\altaffilmark{6},
Avi Shporer\altaffilmark{1},
David Charbonneau\altaffilmark{6},
Jonathan Irwin\altaffilmark{6},
Kristo Ment\altaffilmark{6},
Jennifer G.\ Winters\altaffilmark{6},
Karen A.\ Collins\altaffilmark{6},
Phil Evans\altaffilmark{7},
Tianjun Gan\altaffilmark{8},
Rhodes Hart\altaffilmark{9},
Eric L.N.\ Jensen\altaffilmark{10},
John Kielkopf\altaffilmark{11},
Shude Mao\altaffilmark{8},
William Waalkes\altaffilmark{13}, 
Fran\c cois Bouchy\altaffilmark{12},
Maxime Marmier\altaffilmark{12},
Louise D. Nielsen\altaffilmark{12},
Ga\"el Ottoni\altaffilmark{12},
Francesco Pepe\altaffilmark{12},
Damien S\'egransan \altaffilmark{12},
St\'ephane Udry\altaffilmark{12},
Todd Henry\altaffilmark{20},
Leonardo A. Paredes\altaffilmark{18},
Hodari-Sadiki James\altaffilmark{18},
Rodrigo H.\ Hinojosa\altaffilmark{19},
Michele L.\ Silverstein\altaffilmark{18},
Enric Palle\altaffilmark{21},
Zachory Berta-Thompson\altaffilmark{13},
Misty D.\ Davies\altaffilmark{4},
Michael Fausnaugh\altaffilmark{1},
Ana W.\ Glidden\altaffilmark{1},
Joshua Pepper\altaffilmark{14},
Edward H.\ Morgan\altaffilmark{1},
Mark Rose\altaffilmark{15},
Joseph D.\ Twicken\altaffilmark{16},
Jesus Noel S.\ Villase\~nor\altaffilmark{1},
and the TESS Team
}

\altaffiltext{1}{Department of Physics and Kavli Institute for Astrophysics and Space Science, Massachusetts Institute of Technology, Cambridge, MA 02139, USA}
\altaffiltext{2}{Juan Carlos Torres Fellow}
\altaffiltext{3}{Department of Astronomy, The University of Texas at Austin, Austin, TX 78712, USA and NASA Sagan Fellow}
\altaffiltext{4}{NASA Ames Research Center, Moffett Field, CA 94035, USA}
\altaffiltext{5}{Department of Astrophysical Sciences, Princeton University, Princeton, NJ 08544, USA}
\altaffiltext{6}{Harvard-Smithsonian Center for Astrophysics, 60 Garden Street, Cambridge, MA 02138, USA}
\altaffiltext{7}{El Sauce Observatory, Chile}
\altaffiltext{8}{Department of Physics and Tsinghua Centre for Astrophysics, Tsinghua University, Beijing, China}
\altaffiltext{9}{University of Southern Queensland, Toowoomba, Queensland, Australia 4350}
\altaffiltext{10}{Department of Physics and Astronomy, Swarthmore University, Swarthmore, PA 19081, USA}
\altaffiltext{11}{Department of Physics and Astronomy, University of Louisville, Louisville, KY 40292, USA}
\altaffiltext{12}{Observatoire de l'Universit\'e de Gen\`eve, 51 chemin des Maillettes, 1290 Versoix, Switzerland}
\altaffiltext{13}{Department of Astrophysical and Planetary Sciences, University of Colorado, Boulder, CO 80309, USA}
\altaffiltext{14}{Department of Physics, Lehigh University, Bethlehem, PA 18015, USA}
\altaffiltext{15}{Leidos, Inc., Moffett Field, CA 94035,
USA}
\altaffiltext{16}{SETI Institute, Moffett Field, CA 94035,
USA}
\altaffiltext{17}{Department of Earth and Planetary Sciences, Massachusetts Institute of Technology, Cambridge, MA 02139, USA}
\altaffiltext{18}{Department of Physics and Astronomy, Georgia State University, Atlanta, GA 30302-4106}
\altaffiltext{19}{Cerro Tololo Inter-American Observatory, CTIO/AURA Inc., La Serena, Chile}
\altaffiltext{20}{RECONS Institute, Chambersburg, PA, 17201}
\altaffiltext{21}{Instituto de Astrofisica de Canarias, Tenerife, Spain}

\begin{abstract}

Data from the newly-commissioned \textit{Transiting Exoplanet Survey
  Satellite} (\tess) has revealed a ``hot Earth'' around \target, an
M dwarf located 15~pc away.  The planet has a radius of \tessfitRp{}~$R_\oplus$ and orbits the star every 11 hours.  Although the
existence of an atmosphere around such a strongly irradiated planet is
questionable, the star is bright enough ($I=11.9$, $K=9.1$) for this
possibility to be investigated with transit and occultation spectroscopy.
The star's brightness and the planet's short
period will also facilitate the measurement of the planet's mass
through Doppler spectroscopy.

\end{abstract}

\section{Introduction}
\label{sec:intro}

The {\it Transiting Exoplanet Survey Satellite} (\tess\,)
is a NASA Explorer mission that was launched on April
18, 2018.  The mission's primary objective is to discover hundreds of
transiting planets smaller than Neptune, around
stars bright enough for spectroscopic investigations of planetary
masses and atmospheres \citep{ricker:2015}.
Using four 10\,cm refractive CCD
cameras, \tess\ obtains optical images of a rectangular field spanning
2300~square degrees.  The field is changed every 27.4 days (two
spacecraft orbits), allowing the survey to
cover most of the sky in 2 years.
\tess\ is a wider-field, brighter-star
successor to the successful space-based transit surveys CoRoT
\citep{Barge:2008, haywood2014} and {\it Kepler} \citep{boruckikoi}.

Another way in which \tess\ differs from the previous space missions
is that M dwarfs constitute a larger fraction of the stars being searched,
mainly because of a redder observing bandpass
(600--1000~nm).
Compared to solar-type stars, M dwarfs are advantageous for transit
surveys because the signals are larger for a given planet size, and
because the transits of planets in the ``habitable zone'' are
geometrically more likely and repeat more frequently \citep[see, e.g.,][]{gould:2003, Charbonneau:2007, Latham:2012}. We also know that
close-orbiting planets are very common
around M dwarfs, based on results from the {\it Kepler} survey \citep{dc15,muirhead:2015}.
By focusing on nearby M dwarfs, the pioneering ground-based transit surveys
MEarth and TRAPPIST have discovered four of the most remarkable
planetary systems known today:
GJ\,1214 \citep{charbonneau09}, GJ\,1132 \citep{bertathompson2015},
LHS\,1140 \citep{dittmann2017, ment2018}, and TRAPPIST-1 \citep{gillon16}.

Simulations have shown that \tess\ should be capable of detecting hundreds of
planets around nearby M dwarfs \citep{Sullivan:2015, Bouma:2017,
  ballard2018, muirhead2018, barclay:2018, huang:2018}.
Here, we report the first such detection,
based on data from the first month of the survey.
The planet is $1.32\pm 0.02$ times
larger than the Earth, and orbits the M~dwarf \target\ every 11
hours.  The star, located 15~parsecs away, has a mass and radius that
are about 15\% and 19\% of the Sun's values.  The proximity and
brightness of the star make this system a good candidate for follow-up Doppler and atmospheric spectroscopy.

This {\it Letter} is organized as follows. Section \ref{sec:obs}
presents the data from \tess\, along with follow-up observations with
ground-based telescopes. Section \ref{sec:analysis} describes our
method for determining the system parameters. This section also
explains why the transit-like signal is very likely to represent a
true planet and not an eclipsing binary or other types of ``false
positives.''  Section \ref{sec:dis} compares LHS\,3844\,b with the
other known transiting planets, and discusses some possibilities for
follow-up observations.

\section{Observations and Data analysis}
\label{sec:obs}

\subsection{TESS}
\label{sec:tess}

\tess\ observed \target\ between 2018~Jul~25 and Aug~22, in the first
of 26 sectors of the two-year survey.  The star appeared in CCD~2 of
Camera~3.  The CCDs produce images every 2~seconds, which are summed
onboard the spacecraft into images with an effective exposure time of
30 minutes.  In addition, 2-minute images are prepared for subarrays
surrounding pre-selected target stars, which are chosen primarily for
the ease of detecting transiting planets. LHS\,3844 was
prioritized for 2-minute observations on account of its brightness in
the \tess\ bandpass ($T=11.877$), small stellar radius, and relative
isolation from nearby stars
\citep{Stassun:2017,muirhead2018}.

The 2-minute data consist of 11 by 11 pixel subarrays.  They were
reduced with the Science Processing Operations Center (SPOC) pipeline,
originally developed for the {\it Kepler} mission at the NASA
Ames Research Center \citep{Jenkins:2015,Jenkins:2016}.  For \target,
the signal-to-noise ratio of the transit signals was 32.4.  
The 30-minute data were analyzed independently with the MIT Quick 
Look Pipeline (QLP; Huang et al., in prep).  A transit search with the 
Box Least Square algorithm \citep[BLS, ][]{Kovacs2002} led to a 
detection with a signal-to-noise ratio of 31.6.

\begin{figure*}
\includegraphics[width=\linewidth]{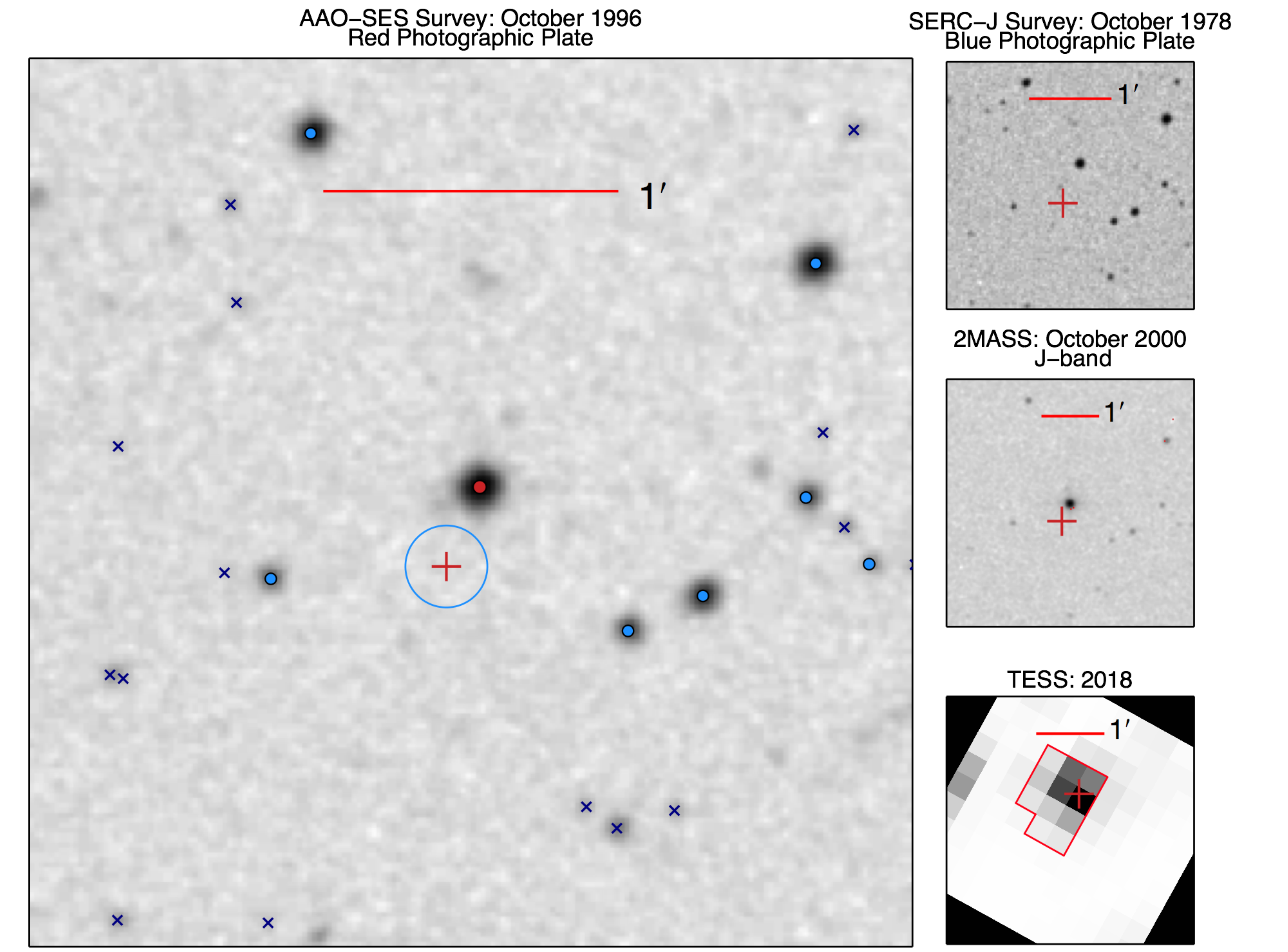}
\caption{Images of the field surrounding LHS 3844. {\it Left.}---From the Anglo-Australian Observatory Second Epoch Survey, obtained
with a red-sensitive photographic emulsion
in 1996.  The red point is the location of \target\ in this image,
and the red cross indicates the current position.
The blue points are stars that are bright enough to potentially
be the source of the transit signal,
while blue crosses are stars that are too faint.
The blue circle shows the $10\sigma$ upper limit
on the motion of the center of light during transits.
The lack of motion rules out the possibility that any
of the surrounding stars is the source of the transit signal.
{\it Upper right.}---From the Science and Engineering Research
Council J survey, obtained with a blue-sensitive photographic
emulsion in 1978.
{\it Middle right.}--From the Two Micron All Sky Survey (2MASS) in $J$-band.
{\it Lower right.}---Summed \tess\ image. }
\label{fig:image}
\end{figure*}

\begin{figure*}
\includegraphics[width=0.95\linewidth]{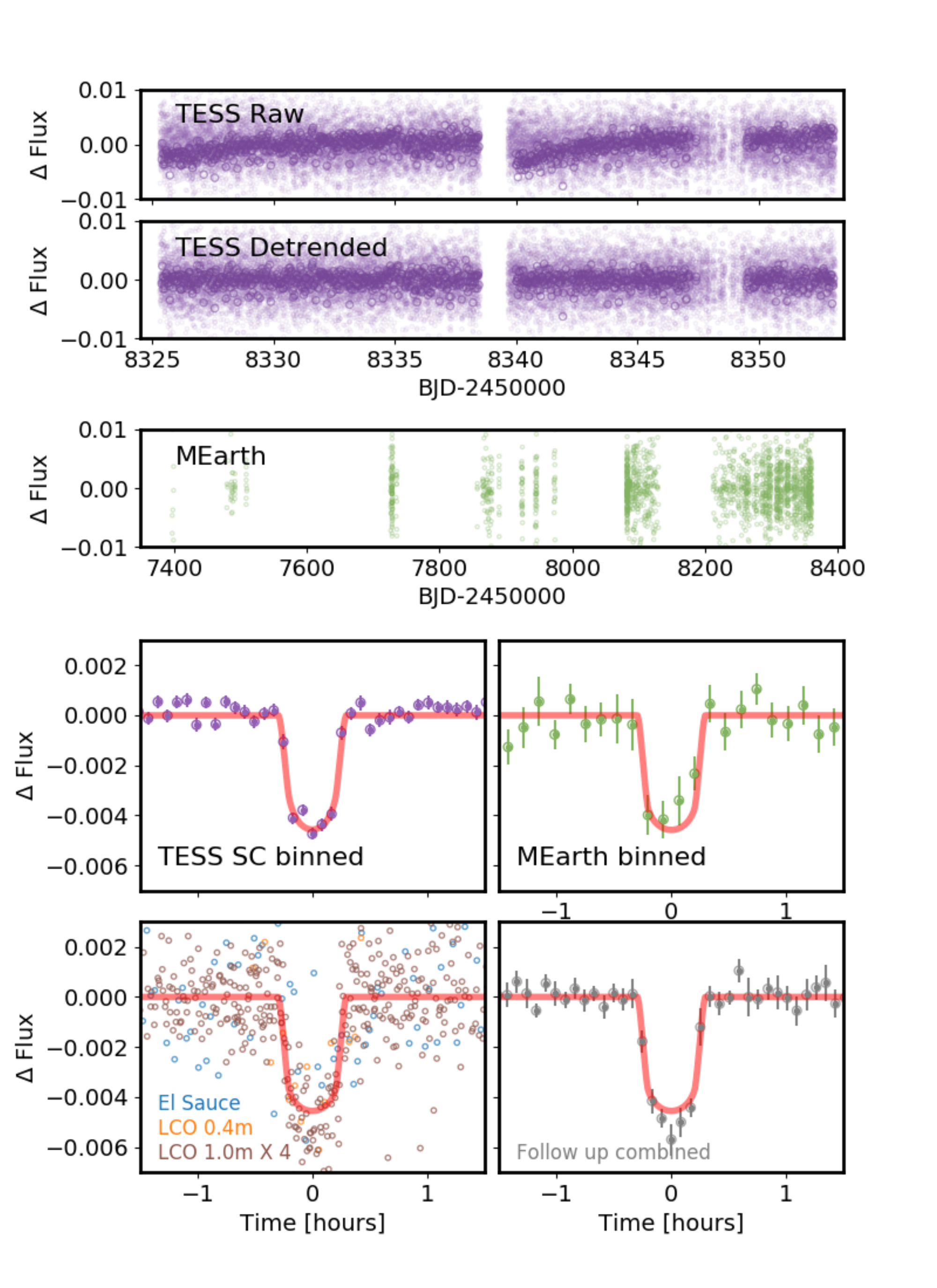}
\vspace{-0.8cm}
\caption{Light curves of \target.
  The top two panels show the \tess\ data, before and after high-pass filtering.
  The middle panel is from the MEarth Observatory, after correcting
  for systematics. The lower grid of four panels are phase-folded
  light curves, along with the best-fitting transit model. The \tess\ data
  points with error bars represent 5\,min averages. The MEarth data points represent 8\,min averages. 
  The bottom two panels show data
  from the \tess\ Follow-up Observing Program, both the original
  data and 5\,min averages.
\label{fig:lc}}
\end{figure*}

For subsequent analysis, we used the 2-minute Pre-search Data Conditioning
light curve from the SPOC pipeline \citep{stumpe},
which was extracted from the photometric aperture depicted in the lower
right panel of Figure~\ref{fig:image}. The resulting light curve is shown
in the top panel of Figure~\ref{fig:lc}. 
To filter out low-frequency
variations, we fitted a basis spline to the 
light curve (excluding the transits and 3$\sigma$ outliers) and 
divided the light curve by the best-fit spline. The result
is shown in the second panel of Figure \ref{fig:lc}.
The interruption in the middle of the time
series occurred when the spacecraft was at perigee, when it reorients
and downlinks the data.  There was also a 2-day interval when the data were
compromised by abnormally unstable spacecraft pointing.  In addition,
we omitted the data collected in the vicinity of ``momentum
dumps,'' when the thrusters are used to reduce the speed of the spacecraft reaction wheels. These
lasted 10--15 minutes and took place every 2.5 days.

\subsection{Ground-based Photometry}
\label{sec:groundphot}

\target~was observed by the ground-based MEarth-South telescope array
as part of normal survey operations \citep{Irwin2015,Dittmann2017a}.
A total of 1935 photometric observations were made between 2016~Jan~10 and 2018~Aug~25.
No transits had been detected prior to the \tess\ detection,
but when the data were revisited,
a BLS search identified a signal with a period and
amplitude consistent with the \tess\ signal (Figure \ref{fig:lc}).
The MEarth data also reveal the stellar rotation period to be 128
days, based on the method described by
\citet{Newton:2016,Newton:2018}.\footnote{Although \citet{Newton:2018} did
  not detect rotational modulation, subsequent data have
  allowed for a ``Grade B'' detection, in the rating system
  described in that work.}

\begin{figure*}
\includegraphics[width=\linewidth]{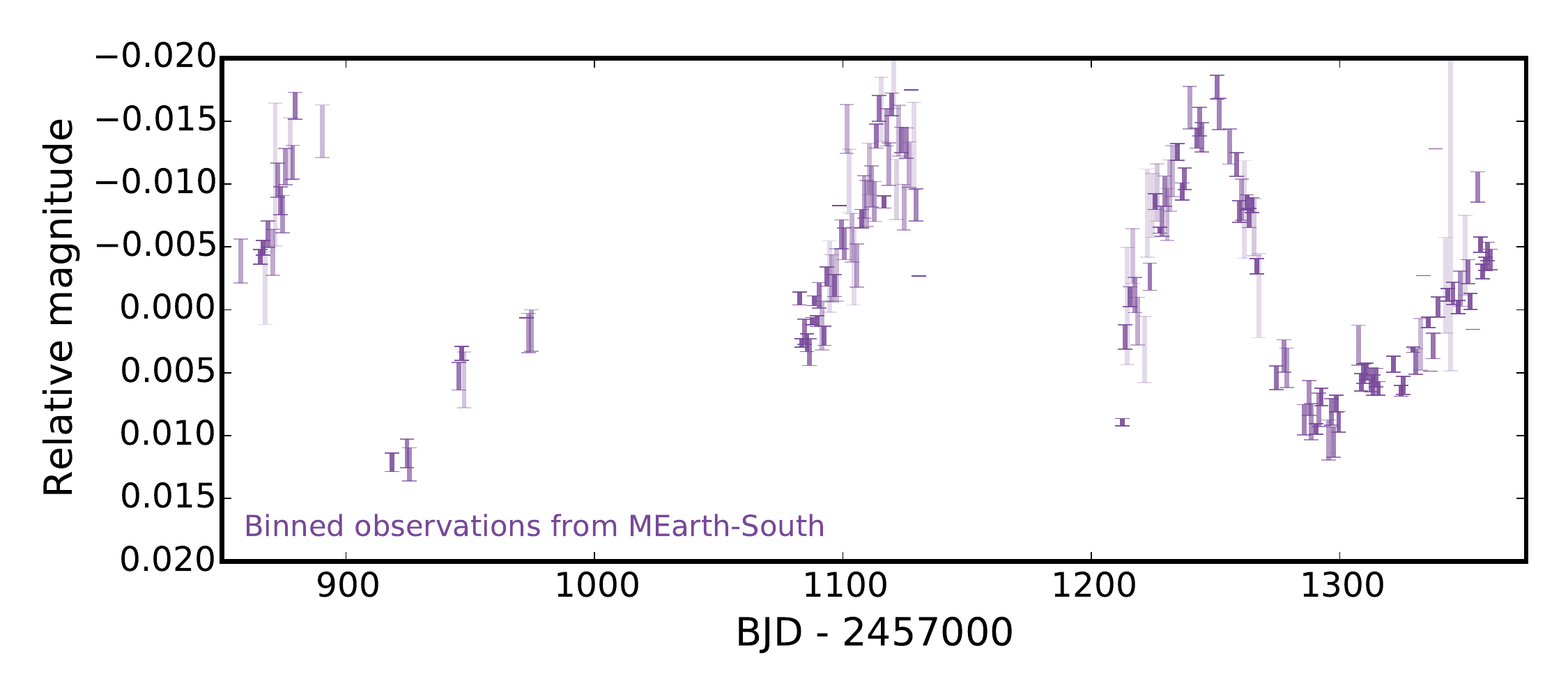}
\caption{Long-term photometric monitoring of \target\ by the MEarth
  Observatory. One-day averages are plotted, with error bars representing
  the standard error of the mean. Data points with higher precision
  are plotted with more opaque symbols. Based on the sinusoidal modulation
  observed in the most recent data, the rotation period is approximately
  $128$ days.}
\label{fig:rotation}
\end{figure*}

Additional ground-based transit observations
were performed as part of the \tess\ Follow-up Observing
Program (TFOP). A full transit was observed on UT 2018~Sep~06 in the $I_C$
band, using the El Sauce Observatory Planewave CDK14 telescope located
in El Sauce, Chile. Five more transits were observed in the Sloan
$i'$ band using telescopes at the Cerro Tololo International
Observatory (CTIO) node of the Las Cumbres Observatory.$\footnote{https://lco.global}$ The transit of UT 2018~Sep~08
was observed with a 0.4\,m telescope, and the transits of
UT~2018~Sep~08, 09, 10 and 16 were observed with a 1.0\,m telescope. The
data are shown in the lower panels of Figure~\ref{fig:lc}.
Together, they confirm the fading events are occurring
and localize the source to within $2\arcsec$ of \target.

\subsection{High-Resolution Spectroscopy} 
\label{sec:spec}

\begin{deluxetable}{lrrl}
\tablewidth{0pc}
\tablecaption{
  Radial Velocities of \target 
  \label{tab:rvs}
}
\tablehead{
  \colhead{BJD} & \colhead{RV} & \colhead{$\sigma_{RV}$} & \colhead{Instrument} \\
  \colhead{} & \colhead{(\kms)} & \colhead{(\kms)} & \colhead{}
}
\startdata
$2458287.9183$   & $-10.626$ & $0.056$ & CHIRON \\
$2458287.9393$   & $-10.667$ & $0.108$ & CHIRON \\
$2458369.6266$   & $-10.719$ & $0.031$ & CHIRON \\
$2458369.6476$   & $-10.732$ & $0.046$ & CHIRON \\
$2458369.6686$   & $-10.724$ & $0.048$ & CHIRON \\
$2458371.653268$ & $-10.600$ & $0.090$ & CORALIE \\
$2458372.780028$ & $-10.540$ & $0.068$ & CORALIE \\ [-2ex]
\enddata
\tablecomments{The 100~\ms\ difference between the average CHIRON and CORALIE velocities
is due to differences in the instrumental zero points, and is not indicative of velocity variation.}
\end{deluxetable}

We obtained optical spectra on UT~2018~Jun~18 and Sep~08 using the
CTIO HIgh ResolutiON (CHIRON) spectrograph on the 1.5\,m telescope of
the CTIO Small and Moderate Aperture Research Telescope System \citep{Tokovinin+2013}. We used the image slicer mode, giving a resolution of about 80{,}000.
The first observation was a pair of 30\,min exposures centered at an
orbital phase of $0.355$. The second observation comprised three
30\,min exposures centered at phase $0.880$. The data were analyzed as
described by \citet{Winters(2018a)}, using a spectrum of Barnard's
Star as a template. The spectra show no evidence of additional lines
from a stellar companion, no sign of rotational broadening, no
detectable H$\alpha$ emission, and no radial velocity variation.

Additional spectroscopy was performed with the CORALIE spectrograph
\citep{queloz:2000,pepe:2017} on the Swiss Euler 1.2\,m telescope at
La Silla Observatory in Chile.  Spectra were obtained on
UT~2018~Sep~10 and 11, at phases 0.211 and 0.645, near the expected
radial-velocity extrema. Radial-velocity calibration was performed
with a Fabry-P\'erot device. With exposure times of 45 and 60\,min,
the signal-to-noise ratio per pixel was about 3 in the vicinity of 600~nm.
Cross-correlations were performed with a weighted M2 binary mask from
which telluric and interstellar lines were removed
\citep{pepe:2002}. Only a single peak was detected. The difference in
radial velocities was $60 \pm 110$~\ms, i.e., not statistically
significant.

To place an upper limit on the radial velocity variation using both
datasets (see Table~\ref{tab:rvs}),
we fitted for the amplitude of a sinusoidal function with
a period and phase specified by the \tess\ transit signal.  The free
parameters were the amplitude $K$, and two additive constants
representing the zeropoints of the CHIRON and CORALIE velocity scales.
The result was $K = -28^{+64}_{-60}\ \ms$, which can be interpreted
as a 3$\sigma$ upper limit of 0.96~\mjup\ on the mass of the
transiting object.

\section{Analysis}
\label{sec:analysis}
\subsection{Stellar parameters}
\label{sec:stellar_parameters}

Using an empirical relationship between mass and $K_s$-band absolute
magnitude \citep{Benedict:2016}, and the parallax from Data Release 2
of the {\it Gaia} mission \citep{GaiaDR2:2018, Lindegren:2018}, 
the mass of \target\ is $0.151 \pm 0.014$ \msun.
The uncertainty 
is dominated by the scatter in the mass-$K_s$ relation.  Based on this 
mass determination, and the
empirical mass-radius relationship of \citet{Boyajian:2012}, the
stellar radius is $0.188\pm0.006$ \rsun.  These results are consistent
with the empirical relationship between radius and absolute $K_s$ magnitude
presented by
\citet{Mann:2015}, which gives $0.189 \pm 0.004$~\rsun.  The
bolometric luminosity is $(2.72\pm0.04)\times 10^{-3}$~\lsun, based on
the observed $V$ and $J$ magnitudes and the bolometric correction from
Table 3 of \citet{Mann:2015}.  Based on these determinations of
$R_\star$ and $L_\star$, the Stefan-Boltzmann law gives an effective
temperature of $3036 \pm 77$~K. The spectral type is M4.5 or M5, based
on a comparison with MEarth survey stars of known spectral types on a
color-magnitude diagram ({\it Gaia}~$G$ versus $H-K_s$).

\subsection{Light curve modeling} 
\label{sec:lcmodeling}

We jointly analyzed the light curves from \tess, MEarth, and TFOP,
using the formalism of
\citet{MandelAgol:2002} as implemented by \citet{Kreidberg(2015)}.  We
assumed the orbit to be circular, and placed Gaussian prior
constraints on the two parameters of a quadratic limb-darkening law
\citep[$u_1 = 0.145\pm0.05$ and $u_2 = 0.54\pm0.05$; ][]{Claret:2018}.
Because of the differing bandpasses of \tess, MEarth, and the TFOP instruments, 
each dataset was allowed to have different values for the limb-darkening
parameters.
We imposed a prior constraint on the
mean stellar density based on the results of
Section~\ref{sec:stellar_parameters}. The model was evaluated with
0.4~min sampling and averaged as appropriate before comparing with the
data.  We used the {\tt emcee} Monte Carlo Markov Chain code of
\citet{ForemanMackey:2012} to determine the posterior distributions
for all the model parameters.  The results are given in
Table~\ref{tab:stellar}.  Figure~\ref{fig:lc} shows the best-fitting
model.

We also performed a fit to the \tess\ data only, without any prior
constraint on the mean stellar density, in order to allow for a
consistency check between the two density determinations. 
Based on the stellar parameters derived in Section~\ref{sec:stellar_parameters}, the mean density is $31.36\pm0.23$ \gcmc,
while the light-curve solution gives $30.0_{-2.8}^{+7.4}$ \gcmc. 
The agreement between these two results
is a sign that the transit signal is from
a planet, and is not an astrophysical false positive.
A related point is that the ratio $\tau/T$ between the
ingress/egress duration and the total duration is
$0.11^{+0.01}_{-0.02}$, and less than 0.14 with 99\% confidence.  This information is used in Section
\ref{sec:lcmodeling} to help rule out the possibility that the fading
events are from an unresolved eclipsing binary.  In general, $\tau/T
\geq R_p/R_\star$, even when the photometric signal includes the
constant light from an unresolved star \citep{morris2018}.

\subsection{Photocenter motion} 
\label{sec:centroids}

Many transit-like signals turn out to be from eclipsing binaries that
are nearly along the same line of sight as the intended target star,
such that the light from the binary is blended together with the
constant light of the intended target star.  These cases can often be
recognized by measuring any motion of the center of light
(``centroid'') associated with the fading events \citep{Wu:2010}.
To do so, we modeled the time series of the $X$ and $Y$ coordinates
of the center of light as though they were light curves,
after removing long-timescale trends by fitting out a cubic spline. 
Based on the fitted depths of the ``centroid transits''
we were able to put $3\,\sigma$ upper limits on centroid shifts
of $\Delta X < 2\times10^{-4}$ and $\Delta Y < 6\times10^{-4}$ pixels,
corresponding to 4.4 and 13.2\,mas. Thus, there is no evidence
for photocenter motion.

\subsection{Possible False Positives}

As mentioned previously, not all transit-like signals are from
transiting planets. Below, we consider the usual alternatives to a
transiting planet, and explain how the available data render them very
unlikely.
\begin{enumerate}
\item {\it The signal is an instrumental artifact.} This is ruled out
  by the detection of the transit signals with ground-based
  telescopes (Section~\ref{fig:lc}).

\item {\it LHS\,3844 is an eclipsing binary star.} This is ruled out
  by the upper limit on radial-velocity variations, corresponding
  to a secondary mass of 0.96~\mjup\ (Section~\ref{sec:spec}).
  In addition, the absence of detectable phase variations 
  in the \tess\ light curve requires that any companion
  be sub-stellar. An 80\mjup\ companion would have produced
  ellipsoidal variations of order 0.1\% \citep[see, e.g.,][]{Shporer:2017},
  which can be excluded.
  
\item {\it Light from a distant eclipsing binary, or a distant star
  with a transiting planet, is blended with that of LHS\,3844.}  We
  can rule out this possibility thanks to the star's high proper
  motion (800~mas~yr$^{-1}$). Because the star moves
  quickly relative to background stars, images from previous
  wide-field surveys allow us to check for faint stars along the
  current line of sight. No sources are detected within 6~mag of
  LHS\,3844, the brightness level that would be required to produce
  0.4\% flux dips. In addition, the ground-based
  observations require the fading source to be within $2\arcsec$ of
  LHS\,3844 (Section~\ref{sec:groundphot}), and the \tess\ images
  reveal no detectable motion of the stellar image during transits
  (Section~\ref{sec:centroids}).
    
\item {\it LHS\,3844 is physically associated with an eclipsing binary
  star.}  The light-curve analysis (Section~\ref{sec:lcmodeling}) requires the
  eclipsing object to be smaller than 14\% of the size of the eclipsed
  star. Since the spectrum is that of an M4-5 dwarf, any secondary
  star would need to be of that size or smaller, implying that the
  eclipsing object is smaller than $0.14 \times 0.15 \rsun$ or
  2.3~$\rearth$. This rules out a stellar binary.
  
\item {\it LHS\,3844 is a binary star and the transiting planet is
  around the secondary star.} The companion would have to be faint and 
  close to LHS\,3844 in order to escape detection by \textit{Gaia} 
  \citep{ziegler, rizzuto}.  Another indication that any secondary star
  needs to be faint is that the {\it Gaia} parallax and apparent
  magnitude are consistent with the properties of a single M dwarf.
  However, if the transiting planet is around such a faint companion, then the
  true transit depth must be less than about 2\% in order for the transiting 
  object to be smaller than about 14\% the size of the eclipsed star. Thus, in order 
  to produce the 0.4\% transit we observe, a secondary star would have to 
  contribute at least 20\% of the total flux in the TESS aperture while
  still escaping detection by \textit{Gaia} and seeing-limited imaging.

\end{enumerate}

Thus, almost all of these scenarios are ruled out, except for the
possibility that the planet is actually orbiting a low-luminosity secondary star.  This scenario seems
contrived, and is {\it a priori} unlikely because of the low
companion fraction for mid M dwarfs \citep{Winters(2018a)}, the small
parameter space for companions which could produce the transits we observe, 
the lower probability for an M dwarf to host a larger
planet compared to a 1.3~$R_\oplus$ planet \citep{berta2013, mulders},
and the tendency of selection effects to favor finding
transits around primary stars \citep{Bouma+2018}.
Probably the only way to rule out this possibility, or
more exotic scenarios, is through precise Doppler monitoring.

\begin{figure*}[ht!]
\vspace{15pt}
\includegraphics[width=\linewidth]{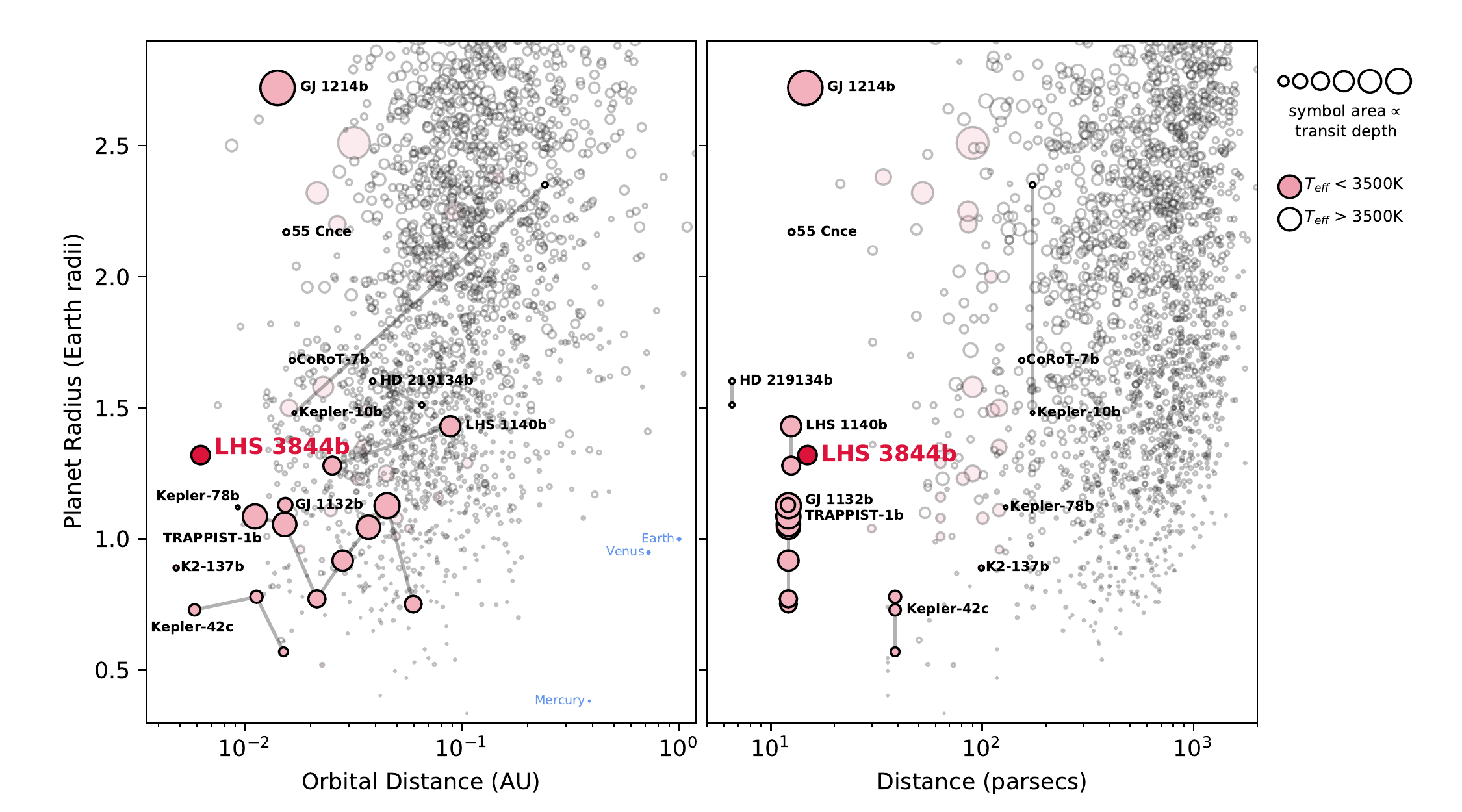}
\caption{
\target\ b in the context of other known exoplanets. 
\textit{Left.}---Planet radius and orbital distance for confirmed
transiting planets. 
\textit{Right.}---Planet radius and distance from Earth.
The area of each circle is proportional to the transit depth of the planet. Pink circles represent host stars with \teff $<3500$~K. Planets in the same system are connected by lines. 
Based on data from the NASA Exoplanet Archive, accessed on 13 September 2018\footnote{\url{https://exoplanetarchive.ipac.caltech.edu/cgi-bin/TblView/nph-tblView?app=ExoTbls\&config=planets}}.
\label{zachstyle}}
\end{figure*}

\section{Discussion}
\label{sec:dis}

\planet\ is one of the closest known planets, both in terms of its distance
from the Earth, and its distance from its host star (see Figure \ref{zachstyle}).
It joins the small club of transiting planets around the Sun's
nearest M dwarf neighbors, which also includes
GJ\,1214\,b \citep{charbonneau09}, GJ\,1132\,b \citep{bertathompson2015},
TRAPPIST-1\,b-h \citep{gillon16}, and LHS\,1140 b-c \citep{dittmann2017, ment2018}.
\planet\ is also the most easily studied example of an ultra-short-period (USP) planet, defined by the simple criterion $P<1$~day \citep{SanchisOjeda+2014,Winn+2018}.
It has the largest transit depth of any known sub-Jovian USP planet,
and is brighter and closer to Earth than the
other well-known systems CoRoT-7 \citep{leger2009}, Kepler-10 \citep{Batalha+2011}, Kepler-42 \citep{Muirhead+2012}, and Kepler-78 \citep{SanchisOjeda+2013}.

As such, \planet\ provides interesting opportunities for atmospheric
characterization through transit and occultation (secondary eclipse) spectroscopy. With an equilibrium temperature
of about 805\,K, and an orbital distance amounting
to only 7.1 stellar radii, it is unclear what type of atmosphere
the planet might have, if any.
If the planet formed at or near this location,
its primordial atmosphere could have been
completely stripped away during the host star's
youth, when it was much more luminous
and chromospherically active.
The observed radius function of the short-period
{\it Kepler} planets
has a dip at around 1.8~$R_\oplus$ that has been
interpreted as a consequence of atmospheric loss.
Planets smaller than 1.8~$R_\oplus$ seem to have
lost their primordial hydrogen-helium atmospheres
due to photoevaporation
\citep{Fulton+2017,LopezFortney2013,OwenWu2013}.
With a radius of 1.32~$R_\oplus$,
we might expect \planet\ to have suffered this process,
too.  In this case, transit spectroscopy would show no variation
in the planetary radius with wavelength, although occultation
spectroscopy could still be used to measure the emission
spectrum of the planet's surface.

Indeed, of all the known planets smaller than 2~$R_\oplus$,
\planet\ has perhaps the most readily detectable
occultations. This is based on a
ranking of the 907 planets in the
NASA Exoplanet Archive by a crude signal-to-noise metric,
\begin{equation}
{\rm S/N} \propto \sqrt{F_\star}~\frac{R_p^2 T_p}{R_\star^2 T_\star},
\end{equation}
which assumes that the star and planet are both radiating
as blackbodies in the Rayleigh-Jeans limit.
Here, $F_\star$ is the star's $K$-band flux,
$T_p$ is the planet's equilibrium temperature,
and $T_\star$ is the star's effective temperature\footnote{See also work by \citet{kempton2018}}. 
According to this metric, \planet\ ranks second, closely trailing 
HD 219134 b, which orbits a much brighter, but larger, star. 
Even then, \planet\ will likely be easier to observe than HD 219134 b thanks 
to its significantly deeper secondary eclipse, which should avoid observational systematic noise floors,
and the planet's ultra-short period, which simplifies scheduling observations.

The ultra-short period will also facilitate the measurement
of the planet's mass through Doppler spectroscopy.  Short periods
lead to stronger signals: assuming
the planet's mass is 2.8~$M_\oplus$, as it would be for
a terrestrial composition, the expected semiamplitude of the
Doppler signal is 8~\ms, which is unusually high for a
rocky planet.  The orbital period is short enough
for the signal to be measured in its entirety in just a few nights.
The orbital period is also
280 times shorter than the stellar rotation period,
allowing for a clear separation of timescales between
the orbital motion and any spurious Doppler signals related
to stellar activity.  

The discovery of a terrestrial planet around a
nearby M dwarf during the first \tess\ observing sector
suggests that the prospects for future
discoveries are bright.  It is worth remembering that 90\% of the
sky has not yet been surveyed by either \tess\ or \Kepler.


\acknowledgments
This work makes use of observations from the LCOGT network. Work by JNW was partially supported by the Heising-Simons Foundation.
We acknowledge the use of \tess\ Alert data, which is currently in a beta test phase, from the \tess\ Science Office. Funding for the \tess\ mission is provided by NASA's Science Mission directorate.
The MEarth team acknowledges funding from the David and Lucile Packard Fellowship for Science and Engineering (awarded to D.C.). This material is based on work supported by the National Science Foundation under grants AST-0807690, AST-1109468, AST-1004488 (Alan T. Waterman Award) and AST-1616624. Acquisition of the CHIRON data was made possible through the support of a grant from the John Templeton Foundation. The opinions expressed in this publication are those of the authors and do not necessarily reflect the views of the John Templeton Foundation. J.A.D. acknowledges support by the Heising-Simons Foundation as a 51 Pegasi b postdoctoral fellow. E.R.N. is supported by an NSF Astronomy and Astrophysics Postdoctoral Fellowship under award AST-1602597.
We thank the Geneva University and the Swiss National Science Foundation for their continuous support for the Euler telescope. 
This research has made use of the NASA Exoplanet Archive, which is operated by the California Institute of Technology, under contract with the National Aeronautics and Space Administration under the Exoplanet Exploration Program.
CXH acknowledges support from MIT's Kavli Institute as a Torres postdoctoral fellow.
AV's work was performed under contract with the California Institute of Technology / Jet Propulsion Laboratory funded by NASA through the Sagan Fellowship Program executed by the NASA Exoplanet Science Institute.
{\it Facilities:} 
\facility{TESS}, 
\facility{CTIO:1.5m (CHIRON)},
\facility{Euler1.2m (CORALIE)},
\facility{LCO:0.4m (SBIG)},
\facility{LCO:1.0m (Sinistro)}
\bibliographystyle{apj}
\bibliography{refs}

\begin{deluxetable*}{lcr}
\tablewidth{0pc}
\tabletypesize{\scriptsize}
\tablecaption{
    Stellar and Planet Parameters for \target
    \label{tab:stellar}
}
\tablehead{
    \multicolumn{1}{c}{~~~~~~~~Parameter~~~~~~~~} &
    \multicolumn{1}{c}{Value}                     &
    \multicolumn{1}{c}{Source}    
}
\startdata
\noalign{\vskip -3pt}
\sidehead{Catalog Information}
~~~~R.A. (h:m:s)                      & ~~\,22:41:59.089     & Gaia DR2\\
~~~~Dec. (d:m:s)                      &  -69:10:19.59    & Gaia DR2\\
~~~~Epoch							  & 2015.5           & Gaia DR2 \\
~~~~Parallax (mas)                    & ~~~\,$67.155 \pm 0.051$ & Gaia DR2\\
~~~~$\mu_{ra}$ (mas yr$^{-1}$)        & ~~$334.357 \pm 0.083$   & Gaia DR2 \\
~~~~$\mu_{dec}$ (mas yr$^{-1}$)       & $-726.974 \pm 0.086$ & Gaia DR2\\
~~~~Gaia DR2 ID                       & 6385548541499112448   &  \\
~~~~TIC ID                            & 410153553 & \\
~~~~LHS ID                            & 3844 & \\
~~~~TOI ID                            & 136  & \\

\sidehead{Photometric properties}
~~~~$TESS$ (mag)\dotfill            & 11.877  & TIC V7         \\
~~~~$Gaia$ (mag)\dotfill            & 13.393 & Gaia DR2               \\
~~~~Gaia RP (mag)\dotfill          & 12.052 & Gaia DR2                 \\
~~~~Gaia BP (mag)\dotfill          & 15.451 & Gaia DR2                 \\
~~~~$V_J$ (mag)\dotfill             & 15.26$\pm$0.03   & RECONS\tablenotemark{a} \\
~~~~$R_{KC}$ (mag)\dotfill          & 13.74$\pm$0.02   & RECONS\tablenotemark{a}  \\
~~~~$I_{KC}$ (mag)\dotfill          & 11.88$\pm$0.02   & RECONS\tablenotemark{a}  \\
~~~~$J$ (mag)\dotfill               & 10.046$\pm$0.023 & 2MASS           \\
~~~~$H$ (mag)\dotfill               & ~~9.477 $\pm$0.023& 2MASS           \\
~~~~$K_s$ (mag)\dotfill             & ~~9.145$\pm$0.023  & 2MASS           \\
\sidehead{Derived properties}
~~~~$\mstar$ ($\msun$)\dotfill      &  $0.151\pm $0.014& Parallax +\citet{Benedict:2016}\tablenotemark{b}\\
~~~~$\rstar$ ($\rsun$)\dotfill      & $0.189\pm$0.006 & Parallax +\citet{Mann:2015} \tablenotemark{c}        \\
~~~~$\loggstar$ (cgs)\dotfill       & $5.06\pm0.01$  &  empirical relation + LC \tablenotemark{d}        \\
~~~~$\lstar$ ($\lsun$)\dotfill      & $0.00272\pm0.0004$  & \citet{Mann:2015}     \\
~~~~$\teffstar$ (K)\dotfill        &  $3036\pm77$  & \tablenotemark{3}\\
~~~~$M_V$ (mag)\dotfill &  $14.39\pm$0.02  & Parallax         \\
~~~~$M_K$ (mag)\dotfill &  $8.272\pm$0.015  & Parallax         \\
~~~~Distance (pc)\dotfill           & \tessfitdistance  & Parallax\\
~~~~\rhostar (\gcmc)\dotfill &  \tessfitrhostar & empirical relation + LC \tablenotemark{d} \\

\sidehead{\Lc{} parameters}
~~~$P$ (days)             \dotfill    &  \tessfitP{}            & \\
~~~$T_c$ (${\rm BJD} - 2457000$)    
      \tablenotemark{b}   \dotfill    &   \tessfitTc{}   &          \\
~~~$T_{14}$ (min)
      \tablenotemark{b}   \dotfill    &   \tessfitTdur{} &         \\
~~~$T_{12} = T_{34}$ (min)
      \tablenotemark{b}   \dotfill    &  \tessfitTingress{}  &      \\
~~~$\arstar$              \dotfill    &  \tessfitaor{}    &     \\
~~~$\rpl/\rstar$          \dotfill    &   \tessfitRratio{} &   \\
~~~$b \equiv a \cos i/\rstar$
                          \dotfill    &   \tessfitb{}   &     \\
~~~$i$ (deg)              \dotfill    &  \tessfitinc{}   &    \\

\sidehead{Limb-darkening coefficients}
~~~$c_1$,  MEarth (linear term)   \dotfill    &  $0.13\pm0.03 $  &        \\
~~~$c_2$,  MEarth (quadratic term) \dotfill  &  $0.53\pm0.03$    &       \\
~~~$c_1,TESS$               \dotfill    & $0.52\pm0.04$     &         \\
~~~$c_2,TESS$               \dotfill    &  $0.46\pm0.01$     &     \\
~~~$c_1,i$               \dotfill    & $0.19\pm0.04$     &         \\
~~~$c_2,i$               \dotfill    &  $0.56\pm0.04$     &     \\

\sidehead{Planetary parameters}
~~~$\rpl$ ($\rearth$)       \dotfill    &   \tessfitRp{}   & \\
~~~$a$ (AU)               \dotfill    &   \tessfitsemi{}  &    \\
~~~$T_{\rm eq}$ (K)        \dotfill   &   \tessfitTeq{}        & \\
~~~$\langle F_j \rangle$ ($10^{9}$~\ergscmsq)
                          \dotfill    & \tessfitFinso{}   &  \\ [-1.5ex]
\enddata
\tablenotetext{a}{The optical photometry is from the RECONS survey,
and was measured according to the procedures described in \citet{Jao(2005),Winters(2015)}.}
\tablenotetext{b}{We adopted the error bar based on the scatter
in the empirical relations described by \citet{Benedict:2016}.}
\tablenotetext{c}{We adopted the error bar based on the scatter
in the empirical relations described by \citet{Mann:2015}.}
\tablenotetext{d}{We fitted the transit light curves with a prior constraint
on the stellar density.}
\tablenotetext{e}{The effective temperature was
determined from the bolometric luminosity and the stellar radius.}

\end{deluxetable*}



\end{document}